\documentclass{article}

\usepackage{authblk}

\usepackage{graphicx} 
\usepackage{calc}

\usepackage{hyperref}
\usepackage{amssymb}
\usepackage{placeins}
\usepackage{amsmath}
\usepackage{acro}
\usepackage{tabularx}
\newcolumntype{s}{>{\hsize=.25\hsize}X}
\newcolumntype{t}{>{\hsize=.33\hsize}X}
\usepackage{subcaption}
\usepackage{booktabs}
\usepackage[numbers]{natbib}

\DeclareAcronym{2slbrp}{short   = {\mbox{2S-LBRP}}, long         = {two-step Laplace-Beltrami regularized projection}}
\DeclareAcronym{arap}{short     = {ARAP}, long                   = {as-rigid-as-possible}}
\DeclareAcronym{bdt}{short      = {BDT}, long                    = {bagged decision tree}}
\DeclareAcronym{bfm}{short      = {BFM}, long                    = {Basel face model}}
\DeclareAcronym{cnn}{short      = {CNN}, long                    = {convolutional neural network}}
\DeclareAcronym{cpd}{short      = {CPD}, long                    = {coherent point drift}}
\DeclareAcronym{ci}{short       = {CI}, long                     = {cephalic index}}
\DeclareAcronym{cr}{short       = {CR}, long                     = {cephalic ratio}}
\DeclareAcronym{ct}{short       = {CT}, long                     = {computed tomography}}
\DeclareAcronym{cvai}{short     = {CVAI}, long                   = {cranial vault asymmetry index}}
\DeclareAcronym{dl}{short       = {DL}, long                     = {deep learning}}
\DeclareAcronym{dt}{short       = {DT}, long                     = {decision tree}}
\DeclareAcronym{ecg}{short      = {ECG}, long                    = {electrocardiography}}
\DeclareAcronym{ecgi}{short     = {ECGI}, long                   = {electrocardiographic imaging}}
\DeclareAcronym{fnn}{short      = {FNN}, long                    = {feedforward neural network}}
\DeclareAcronym{gan}{short      = {GAN}, long                    = {generative adversarial network}}
\DeclareAcronym{gpa}{short      = {GPA}, long                    = {generalized Procrustes analysis}}
\DeclareAcronym{gpmm}{short     = {GPMM}, long                   = {Gaussian process morphable model}}
\DeclareAcronym{icp}{short      = {ICP}, long                    = {iterative closest points}}
\DeclareAcronym{icpdlbrp}{short = {\mbox{ICPD-LBRP}}, long       = {iterative coherent point drift with Laplace-Beltrami regularized projection}}
\DeclareAcronym{icpd}{short     = {ICPD}, long                   = {iterative coherent point drift}}
\DeclareAcronym{gpu}{short      = {GPU}, long                    = {graphics processing unit}}
\DeclareAcronym{knn}{short      = {kNN}, long                    = {k-nearest-neighbors}}
\DeclareAcronym{lbrp}{short     = {LBRP}, long                   = {Laplace-Beltrami regularized projection}}
\DeclareAcronym{lb}{short       = {LB}, long                     = {Laplace-Beltrami}}
\DeclareAcronym{lda}{short      = {LDA}, long                    = {linear discriminant analysis}}
\DeclareAcronym{lm}{short       = {LM}, long                     = {landmark}}
\DeclareAcronym{map}{short      = {MAP}, long                    = {maximum a posteriori}}
\DeclareAcronym{mri}{short      = {MRI}, long                    = {magnetic resonance imaging}}
\DeclareAcronym{ml}{short       = {ML}, long                     = {machine learning}}
\DeclareAcronym{mlp}{short      = {MLP}, long                    = {multi layer perceptron}}
\DeclareAcronym{nb}{short       = {NB}, long                     = {naïve Bayes}}
\DeclareAcronym{nicpa}{short    = {\mbox{N-ICP-A}}, long         = {nonrigid iterative closest points affine}}
\DeclareAcronym{nicpt}{short    = {\mbox{N-ICP-T}}, long         = {nonrigid iterative closest point translation}}
\DeclareAcronym{nn}{short       = {NN}, long                     = {neural network}}
\DeclareAcronym{obb}{short      = {OBB}, long                    = {oriented bounding boxes}}
\DeclareAcronym{osnicp}{short   = {\mbox{OS-N-ICP}}, long        = {optimal step nonrigid iterative closest points}}
\DeclareAcronym{pca}{short      = {PCA}, long                    = {principal component analysis}}
\DeclareAcronym{pdm}{short      = {PDM}, long                    = {point distribution model}}
\DeclareAcronym{ppca}{short     = {PPCA}, long                   = {probabilistic principal component analysis}}
\DeclareAcronym{psm}{short      = {PSM}, long                    = {posterior shape model}}
\DeclareAcronym{qda}{short      = {QDA}, long                    = {quadratic discriminant analysis}}
\DeclareAcronym{ransac}{short   = {RANSAC}, long                 = {random sample consensus}}
\DeclareAcronym{rf}{short       = {RF}, long                     = {random forest}}
\DeclareAcronym{rms}{short      = {RMS}, long                    = {root mean squared}}
\DeclareAcronym{shap}{short     = {SHAP}, long                   = {SHapley Additive exPlanations}}
\DeclareAcronym{sids}{short     = {SIDS}, long                   = {sudden infant death syndrome}}
\DeclareAcronym{ssm}{short      = {SSM}, long                    = {statistical shape model}}
\DeclareAcronym{ssim}{short     = {SSIM}, long                   = {structural similarity index measure}}
\DeclareAcronym{ssimcc}{short   = {SSIM\textsubscript{cc}}, long = {structural similarity index measure to closest clinical sample}}
\DeclareAcronym{sto}{short      = {STO}, long                    = {sellion tragion orientation}}
\DeclareAcronym{svd}{short      = {SVD}, long                    = {singular value decomposition}}
\DeclareAcronym{svm}{short      = {SVM}, long                    = {support vector machine}}
\DeclareAcronym{wpca}{short     = {WPCA}, long                   = {weighted principal component analysis}}

\renewcommand*\vec[1]{{\mathbf{\mathrm{#1}}}}
\newcommand*\Space[1]{\mathbb{#1}} 

\begin{document}

\title{Impact of Data Synthesis Strategies for the Classification of Craniosynostosis}
\date{April 2023}

\author[1]{Matthias Schaufelberger}
\author[3]{Reinald Peter Kühle}
\author[1]{Andreas Wachter}
\author[3]{Frederic Weichel}
\author[2]{Niclas Hagen}
\author[2]{Friedemann Ringwald}
\author[2]{Urs Eisenmann}
\author[3]{Jürgen Hoffmann}
\author[3]{Michael Engel}
\author[3]{Christian Freudlsperger}
\author[1]{Werner Nahm}
\affil[1]{Institute of Biomedical Engineering (IBT), Karlsruhe Institute of Technology (KIT), Kaiserstr. 12, 76137 Karlsruhe, Germany}
\affil[2]{Institute of Medical Informatics, Heidelberg University Hospital, Im Neuenheimer Feld 130.3, Heidelberg, Germany}
\affil[3]{Department of Oral, Dental and Maxillofacial Diseases, Heidelberg University Hospital, Im Neuenheimer Feld 400, Heidelberg, Germany}

\maketitle

\section*{Abstract}\label{sec:scAbstract}

\noindent \textit{Introduction:} Photogrammetric surface scans provide a 
radiation-free option to assess and classify craniosynostosis.  Due to the low 
prevalence of craniosynostosis and high patient restrictions, clinical data is 
rare.  Synthetic data could support or even replace clinical data for the 
classification of craniosynostosis, but this has never been studied 
systematically. \\

\noindent \textit{Methods:} We test the combinations of three different 
synthetic data sources: a \ac{ssm}, a \ac{gan}, and image-based \acl{pca} for 
a \ac{cnn}-based classification of craniosynostosis.  The \ac{cnn} is trained 
only on synthetic data, but validated and tested on clinical data.  \\

\noindent \textit{Results:} The combination of a \ac{ssm} and a \ac{gan} 
achieved an accuracy of more than 0.96 and a F1-score of more than 0.95 on the 
unseen test set.  The difference to training on clinical data was smaller than 
0.01.  Including a second image modality improved classification performance 
for all data sources.  \\

\noindent \textit{Conclusion:} Without a single clinical training sample, a 
\ac{cnn} was able to classify head deformities as accurate as if it was 
trained on clinical data.  Using multiple data sources was key for a good 
classification based on synthetic data alone.  Synthetic data might play an 
important future role in the assessment of craniosynostosis.  \\

\acresetall

\section{Introduction}\label{sec:scIntro}

Craniosynostosis is a group of head deformities affecting infants involving 
the irregular closure of one or multiple head sutures and its prevalence is 
estimated to be between four and ten cases per 10,000 live 
births~\citep{french1990}.  As described by Virchow's law~\citep{persing1989}, 
depending on the affected suture distinct types of head deformities arise.  
Genetic mutations have been identified as one of the main causes of 
craniosynostosis~\citep{coussens2007,boulet2008}, which has been linked to 
increased intracranial pressure~\citep{renier1982} and decreased brain 
development~\citep{kappsimon2007}.  The most-performed therapy is surgical 
intervention consisting of resection of the suture and cranial remodeling of 
the skull. It has a high success rate~\citep{fearon2009} and is usually 
performed within the first two years of age.  Early diagnosis is crucial and 
often involves palpation, cephalometric measurements, and medical imaging.  
\Ac{ct} imaging is the gold standard for diagnosis, but makes use of harmful 
ionizing radiation which should be avoided, especially for very young infants.  
Black-bone \ac{mri}~\citep{saarikko2020} is sometimes performed, but requires 
sedation of the infants to impede moving artifacts.  3D photogrammetric 
scanning enables the creation of 3D surface models of the child's head and 
face and is a radiation-free, cost-effective, and fast option to quantify head 
shape.  It can be employed in a pediatrician's office and has potential to be 
used with smartphone-based scanning approaches~\citep{barbero-garcia2020}.

Due to the low prevalence, craniosynostosis is included in the list of rare 
diseases by the American National Organization for Rare Disorders.  Beside the 
few data, strict patient data regulations, and difficulties in anonymization 
(photogrammetric recordings show head and face), there are no publicly 
available clinical datasets of craniosynostosis patients available online.  
Synthetic data could potentially be used as a substitute to develop algorithms 
and approaches for the assessment of craniosynostosis, but only one synthetic 
dataset based on a \ac{ssm} from our group~\citep{schaufelbergermatthias2021} 
has been made publicly available so far.  Scarce training data and high class 
imbalance due to the different prevalences of the different types of 
craniosynostosis~\citep{boulet2008} call for the usage of synthetic data to 
support or even replace clinical datasets as the primary resource for 
\ac{dl}-based assessment and classification.  The inclusion of synthetic data 
could facilitate training due to the reduction of class imbalance and increase 
the classifier's robustness and performance.  Additionally, synthetic data may 
also be used as a cost-effective way to acquire the required training material 
for classification models without manually labeling and exporting a lot of 
clinical data.  Using synthetic data for classification studies in a 
supporting manner or as a full replacement for clinical data has gained 
attraction in several fields of biomedical engineering (e.g.\ 
\citep{nagel2022,sanchez2021}), especially if clinical data is not abundant.  
While classification approaches of craniosynostosis on \ac{ct} 
data~\citep{mendoza2014}, 2D images~\cite{tabatabaei2021}, and 3D 
photogrammetric surface 
scans~\citep{dejong2020,schaufelberger2022a,schaufelberger2023} have been 
proposed, the dataset sizes were below 500 samples (e.g.\ 
\citep{schaufelberger2023}, \citep{dejong2020}, and \citep{mendoza2014}) and 
contained a high class imbalance.  The usage of synthetic data is a 
straightforward way to increase training size and stratify class distribution.

However, although the need for synthetic data had been 
acknowledged~\citep{dejong2020}, synthetic data generation for the 
classification of head deformities has not been systematically explored yet.  
With the scarce availability of clinical data and multiple options of 
synthetic data generation available, we aim to test the effectiveness of 
multiple data synthesis methods both individually and as multi-modal 
approaches for the classification of craniosynostosis.  Using synthetic data 
as training material facilitates not only the development of larger and more 
robust classification approaches, but also makes data sharing easier and 
increases data availability.  A popular approach for 3D data synthesis is 
statistical shape modeling.  It describes the approach to model 3D geometric 
shape variations by means of statistical analysis. With the application of 
head deformities, they have been employed to distinguish clinical head 
parameters~\citep{meulstee2017}, to evaluate head shape 
variations~\citep{rodriguezflorez2017}, to assess therapy 
outcome~\citep{heutinck2021}, and to classify 
craniosynostosis~\citep{schaufelberger2022a}.  Although their value in the 
clinical assessment of craniosynostosis has been shown, the impact of 
\ac{ssm}-based data augmentation for the classification of craniosynostosis 
has not been evaluated yet.  With the introduction of a conversion of the 3D 
head geometry into a 2D image, image-based \ac{cnn}-based 
classification~\citep{schaufelberger2023} can be applied on low-resolution 
images.  \Acp{gan}~\citep{goodfellow2014} have been suggested as a data 
augmentation tool~\citep{dejong2020} and have been able to increase 
classification performance for small datasets~\citep{pinetz2019}.  

The goal of this work is to employ a classifier based on synthetic data, using 
three different types of data synthesis strategies: \ac{ssm}, \ac{gan}, and 
image-based \ac{pca}. The three modalities are systematically compared 
regarding their capability in the classification of craniosynostosis when 
trained only on synthetic data.  We will demonstrate that the classification 
of craniosynostosis is possible with a multi-modal synthetic dataset with a 
similar performance to a classifier trained on clinical data.  Additionally, 
we propose a \ac{gan} design tailored towards the creation of low-resolution 
images for the classification of craniosynostosis.  Both the \ac{gan}, the 
different \acp{ssm}, and \ac{pca}, were made publicly available along as all 
the 2D images from the synthetic training, validation and test sets.  

\section{Methods}\label{sec:scMethods}

\subsection{Dataset and Preprocessing}\label{subsec:datasetPrepro}

All data from this study was provided from the Department of Oral and 
Maxillofacial Surgery of the Heidelberg University Hospital, in which patients 
with craniosynostosis are routinely recorded for therapy planning and 
documentation purposes.  The recording device is a photogrammetric 3D scanner 
(Canfield VECTRA-360-nine-pod system, Canfield Science, Fairfield, NJ, USA).  
We used a standardized protocol which had been examined and approved by the 
Ethics Committee Medical Faculty of the University of Heidelberg (Ethics 
number S-237/2009).  The study was carried out according to the Declaration of 
Helsinki and written informed consent was obtained from parents. 

\begin{figure}[htbp]
\centering
\includegraphics[width=0.6\textwidth]{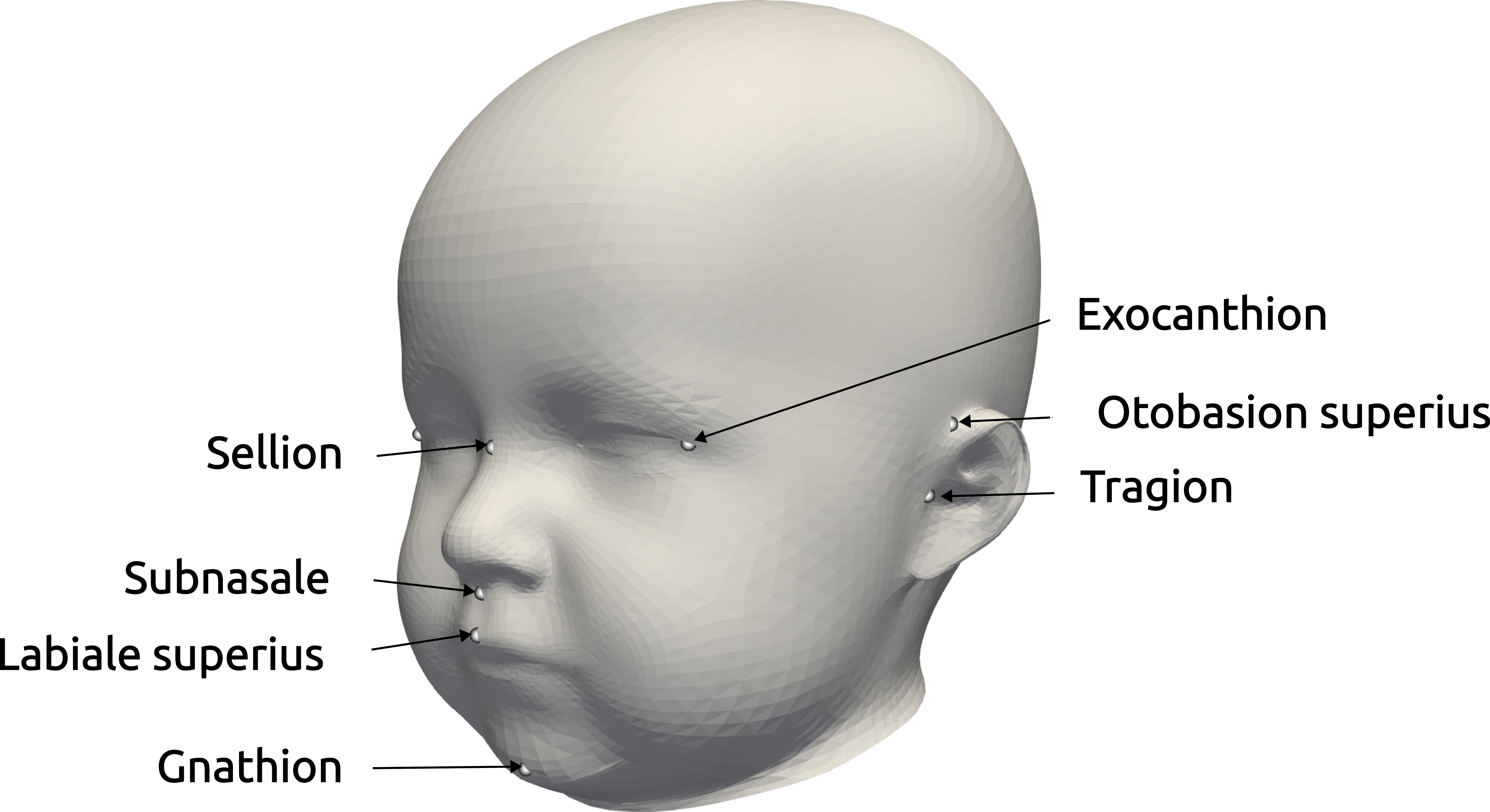}
\caption{Landmarks provided in the dataset, used for the alignment for 
    statistical shape modeling and the coordinate system creation of the 
    distance maps~\cite{schaufelberger2023}.  The three landmarks on the right 
exist for both left and right part of the head.}%
\label{fig:lms}
\end{figure}

\begin{figure}[htbp]
\centering
\includegraphics[width=0.4\textwidth]{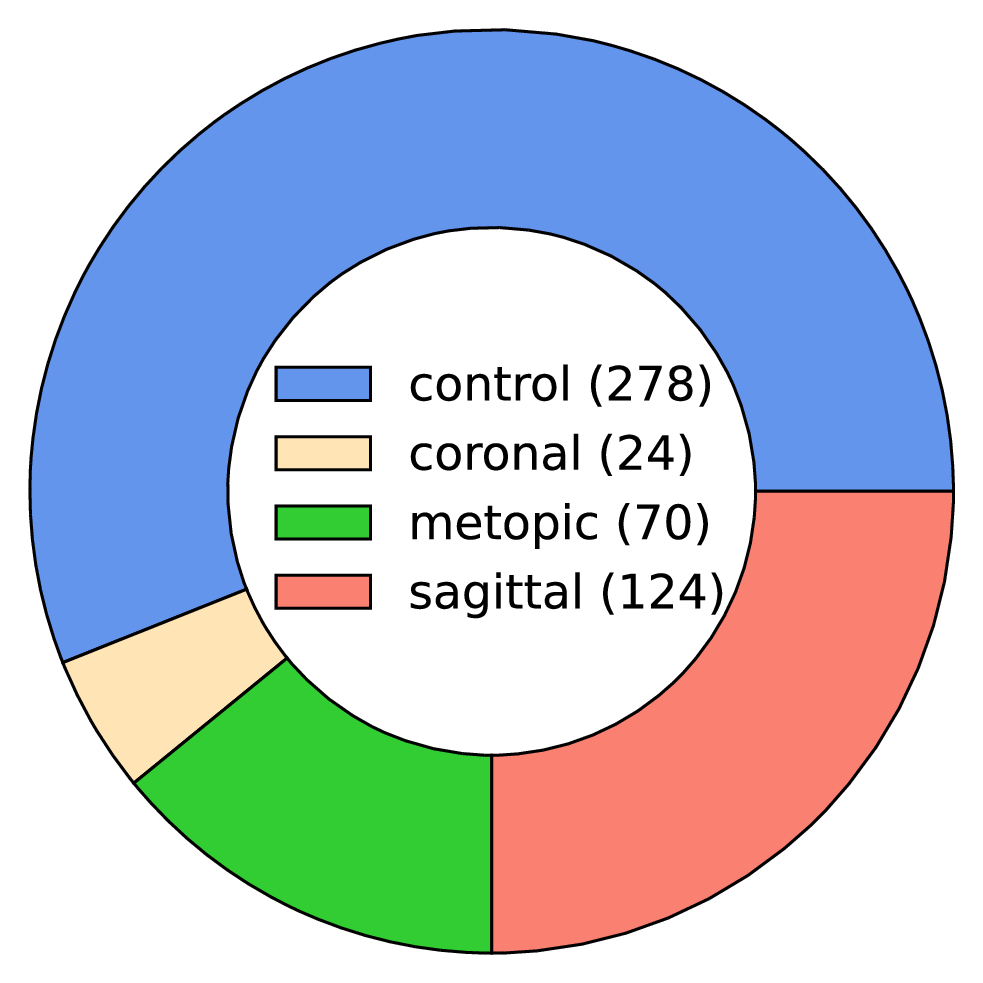}
\caption{Pie chart of the class ratios in the clinical dataset (control 56\,\%, 
coronal 5\,\%, metopic 14\,\%, sagittal 25\,\%).  The legend in the center 
shows the absolute number of samples in the dataset (in total 496 samples).}%
\label{fig:piechart}
\end{figure}

\begin{figure}[hbtp]
\centering
\includegraphics[width=0.8\columnwidth]{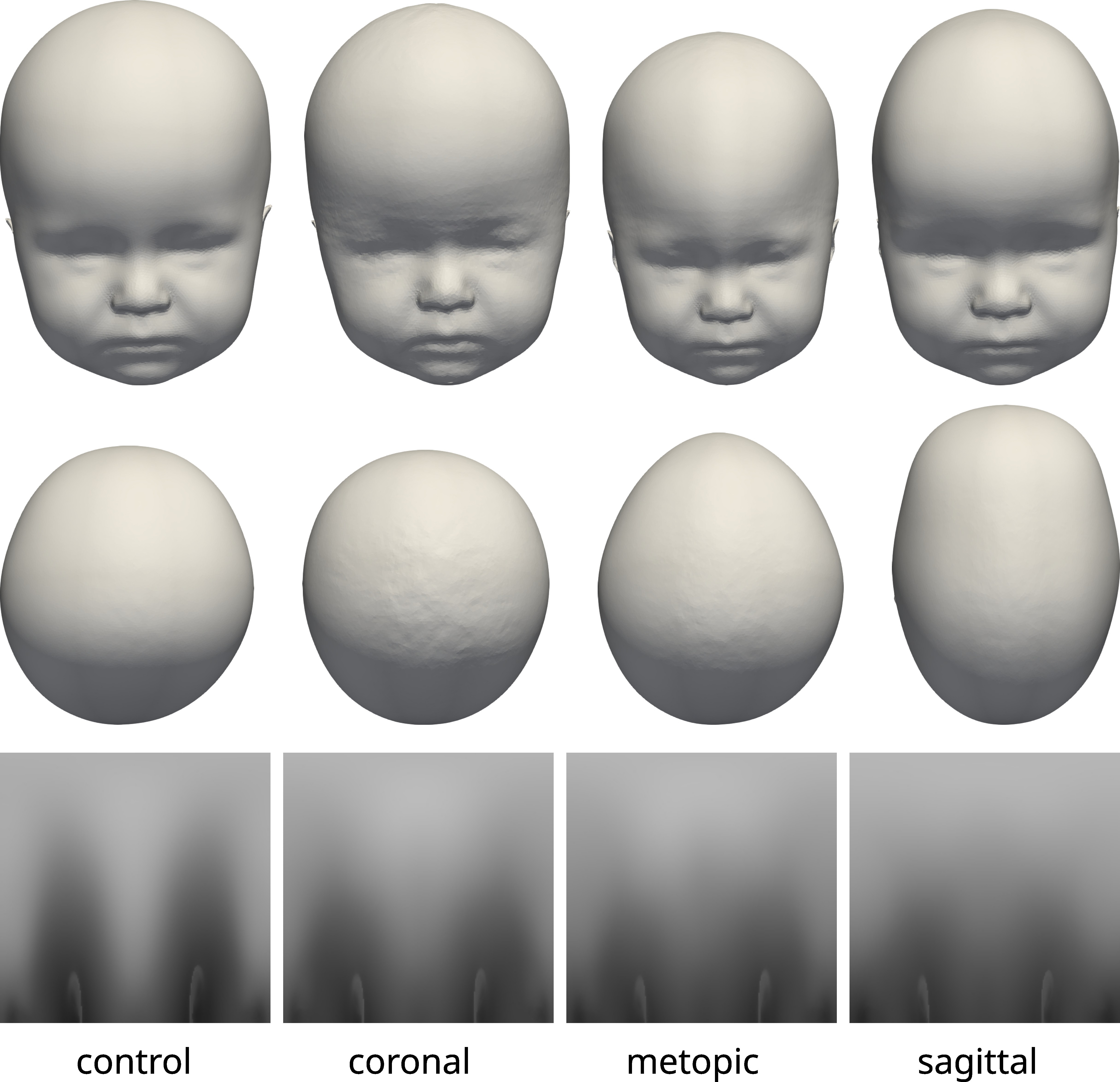}
\caption{The four classes of the dataset with their distinct head shapes and 
    their resulting distance maps representation. Top row: frontal view, middle row: 
    top view, bottom row: 2D distance maps.}\label{fig:classes}
\end{figure}

Each data sample was available as a 3D triangular surface mesh.  We selected 
the 3D photogrammetric surface scans from all available years (2011--2021). If 
multiple scans for the same patient were available, we selected only the last 
preoperative scan to avoid duplicate samples of the same patients.  All 
patient scans had been annotated by medical staff with their diagnosis and 10 
cephalometric landmarks. Fig.~\ref{fig:lms} shows the available landmarks on 
the dataset.  We retrieved patients with coronal suture fusion (brachycephaly 
and unilateral anterior plagiocephaly), sagittal suture fusion 
(scaphocephaly), and metopic suture fusion (trigonocephaly), as well as a 
control group  with the dataset distribution displayed in 
Fig.~\ref{fig:piechart}.  Besides healthy subjects, the control group also 
contained patients suffering from mild positional plagiocephaly without suture 
fusion.  Subjects with positional plagiocephaly in the control group were 
treated with helmet therapy or laying repositioning. In contrast, all patients 
suffering from craniosynostosis required surgical treatment and underwent 
remodeling of the neurocranium.  The four head shape resulting from 
craniosynostosis are visualized in Fig.~\ref{fig:classes}.

We used the open-source Python module \texttt{pymeshlab}~\citep{cignoni2008} 
(version 2022.2) to automatically remove some recording artifacts such as 
duplicated vertices and isolated parts. We also closed holes resulting from 
incorrect scanning and removed irregular edge lengths by using isotropic 
explicit re-meshing~\citep{pietroni2010} with a target edge length of 1\,mm.  
In an earlier work~\citep{schaufelberger2023}, we defined a 2D encoding of the 
3D head shape (``distance maps'', displayed in Fig.~\ref{fig:classes}, bottom 
row) which was also included in the pre-processing pipeline with the default 
parameter of~\citep{schaufelberger2023}.

\begin{figure}[htbp]
\centering
\includegraphics[width=0.8\textwidth]{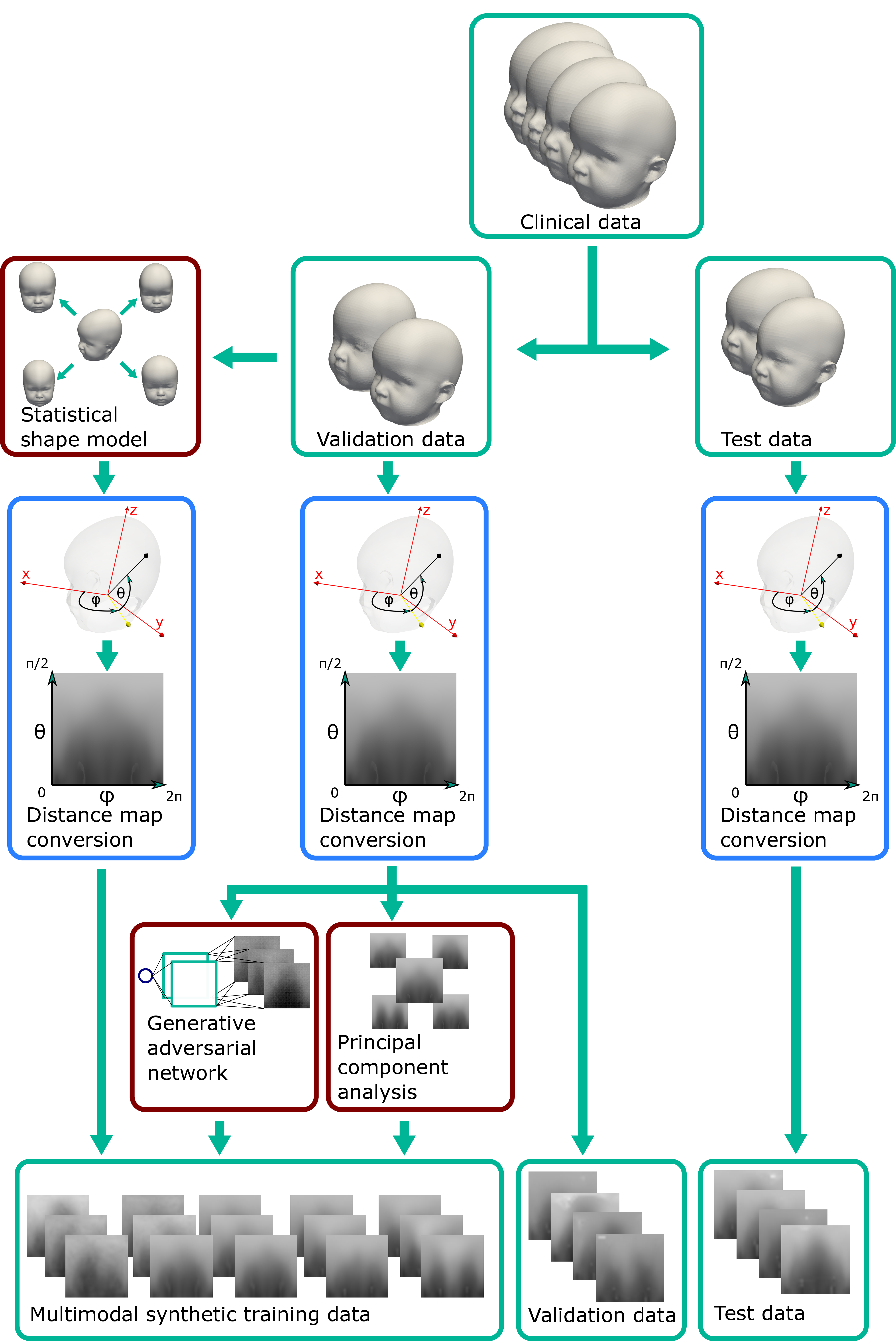}
\caption{Data subdivision for the synthetic-data-based classification and the creation of 
synthetic data.  The test set was separated initially from the dataset, while 
the validation set was used to produce the synthetic samples on which the 
\acs{cnn} was trained.  Green: data, blue: 3D-2D image conversion, dark red: 
generative models.}%
\label{fig:scDataSubdivision}
\end{figure}

\subsection{Data subdivision}\label{subsec:scSchematic}

We did not use the full clinical dataset (validation and test set according to 
Fig.~\ref{fig:scDataSubdivision}) as training data for the data generation 
models (\ac{gan}, \ac{ssm}, and \ac{pca}) since the statistical information of 
the test set would be included in the synthetic data sources, leading to 
leakage (an overestimation of the model performance due to statistical 
information ``leaking'' into the test set).  Instead, we chose the schematic 
displayed in Fig.~\ref{fig:scDataSubdivision}.  We used a stratified 50--50 
split of the clinical data and used one half of the samples as the validation 
set and the other half as the test set.  

The test set was separated from the validation set, only to be used for the 
final evaluation of the classifier.  Following this approach, the test set did 
neither have any influence on the synthetic data, nor was it incorporated in 
the validation set and should therefore be a true representation of unknown 
data to the classifier.  The validation set was used to select the best 
network during training and for hyperparameter tuning, but not as training 
material.  Additionally it was used as the original (training) data on which 
we built the synthetic image generators.  The synthetic training set was then 
created from the validation set according to the three data synthesis 
approaches described below: \ac{ssm}, \ac{gan}, and \ac{pca}.  The three 
approaches operated on different domains: While the \ac{ssm} was applied 
directly on the 3D surface scans, the \ac{gan} and the \ac{pca} used the 2D 
distance map images.  All images were created as 28$\times$28-sized 
craniosynostosis distance maps which was sufficient for good classification in 
an earlier study~\citep{schaufelberger2023}.  We describe each of the three 
individual approaches \ac{ssm}, \ac{gan}, and \ac{pca} below. 

\subsection{Data Synthesis}\label{subsec:scDataSynthesis}

\subsubsection{\Acl{ssm}}

\begin{figure}[htbp]
\centering
\includegraphics[width=\textwidth]{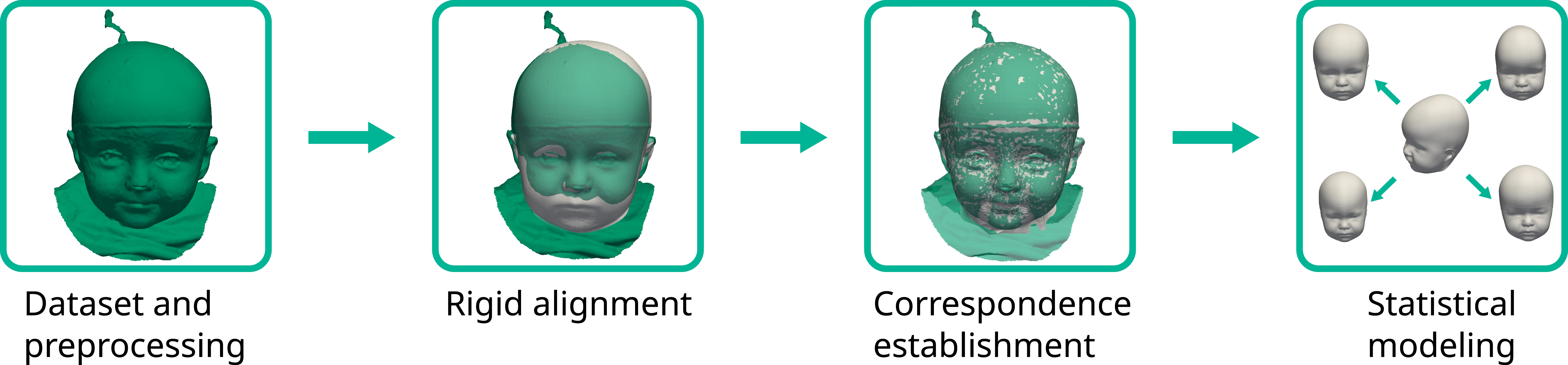} \\
\caption{The statistical shape model pipeline employed in this study.  The 
target scan is colored green with the deforming template in white.}%
\label{fig:fun:schematicSsm}
\end{figure}

The pipeline for the \ac{ssm} creation (similar to~\citep{dai2017}) consists 
of initial alignment, dense correspondence establishment, and statistical 
modeling to extract the mean shape and the principal components from the 
sample covariance matrix (see also Fig.~\ref{fig:fun:schematicSsm}).  For 
correspondence establishment, we employed template morphing.

We used the mean shape of our previously published 
\ac{ssm}~\citep{schaufelbergermatthias2021} as a template which would be 
morphed onto each of the target scans.  Procrustes analysis was employed on the ten 
cephalometric landmarks to obtain a transformation including translation, 
rotation, and isotropic scaling from the template to each target according to 
the cephalometric landmarks on the face and ears.  For correspondence 
establishment, we employed the \ac{lbrp} approach~\citep{dai2019} to morph the 
template onto each of the targets.  We used two iterations: a high stiffness 
fit (providing a now landmark-free transformation from template to the target, 
improving the alignment also from the back of the head not covered with the 
landmarks) and a low stiffness fit (allowing the template to deform very close 
to the targets~\citep{dai2020}).  The deformed templates were then in dense 
correspondence, sharing the same point IDs across all scans and were used for 
further processing. 

\Ac{gpa} was performed to remove both rotational and translational components 
on all the morphed templates so that the mean shape could be determined and 
removed.  The remaining zero mean data matrix served as a basis for the 
principal component analysis.  To counterbalance higher point density in the 
facial regions, we used weighted \ac{pca} instead of ordinary \ac{pca} for the 
statistical modeling.  The weights were assigned according to the surface area 
that each point encapsulated and computed using the area of each triangle of the 
surface model.  We created one \ac{ssm} for each class, ensuring that the 
models were independent from each other and did not contain influences from 
the other classes.  We cut off the coefficient vectors after 95\,\% of the 
normalized variance to remove noise and ensured only the most important 
components were included in the \acp{ssm}.  The synthesis of the model 
instances could then be performed as

\begin{equation}
    \vec{s} = \vec{\bar{s}} + \vec{V} \vec{\Lambda}^\frac{1}{2} \vec{\alpha},
\end{equation}

with $\vec{\bar{s}}$ denoting the mean shape, $\vec{V}$ the principal 
components, $\vec{\Lambda}$ the sample covariance matrix, and $\vec{\alpha}$ the 
shape coefficient vector.  We created 1000 random shapes of each class using a 
Gaussian distribution of the shape coefficient vector and created 
craniosynostosis distance maps for each sample.

\subsubsection{Image-based \acl{pca}}

We used ordinary \ac{pca} as the last modality to generate 2D image data.  
While the \ac{ssm} also made use of \ac{pca} in the 3D domain, image-based 
\ac{pca} operated directly on the 2D images.  This was a computationally 
inexpensive and less sophisticated alternative to both \acp{gan} and \acp{ssm} 
since neither extensive model training and hyperparameter tuning, nor 3D 
morphing and correspondence establishment was required.  We employed ordinary 
\ac{pca} for each of the four classes separately and we again created 1000 
samples for each class.  Since \ac{ssm} is related to \ac{pca}, the data 
synthesis could be performed as

\begin{equation}
    \vec{i} = \vec{\bar{i}} + \vec{V} \vec{\Lambda}^\frac{1}{2} \vec{\alpha},
\end{equation}

with $\vec{\bar{i}}$ denoting the mean image in vectorized shape, $\vec{V}$ 
again the principal components, $\vec{\Lambda}$ the sample covariance matrix, 
and $\vec{\alpha}$ the coefficient vector of the principal components.  We 
again drew 1000 random vectors from a Gaussian distribution and transformed 
them back into 2D image-shape.

\subsubsection{\Acl{gan}}

\begin{figure}[htpb]
\centering
\includegraphics[width=\textwidth]{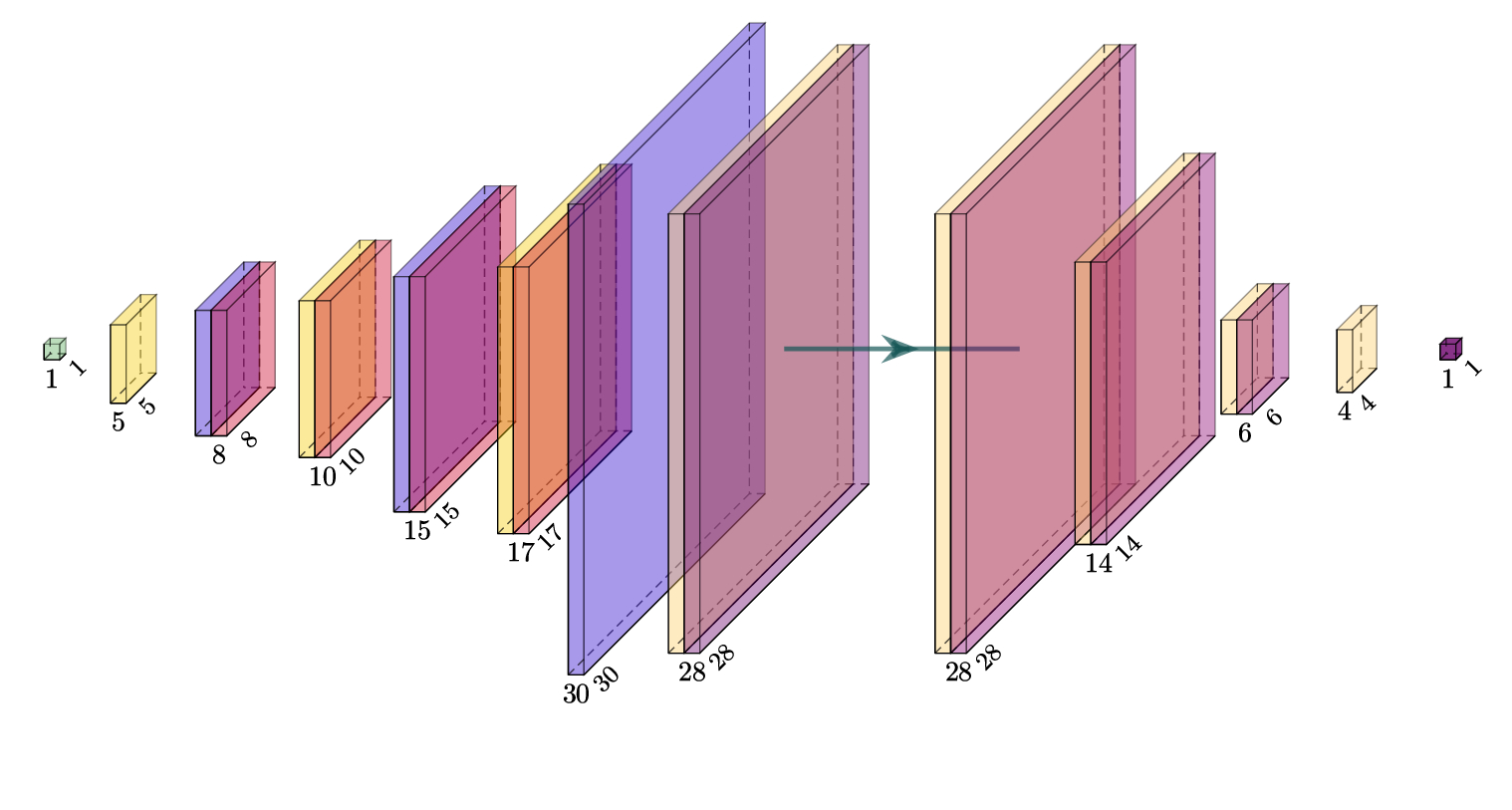}
\caption{Visualization of the intermediate image sizes from the used \acs{gan} 
model.  Left: generator, right: critic (discriminator).  The filter kernel 
sizes are described in the Appendix~\ref{ap:gan}.}%
\label{fig:scGanStructure}
\end{figure}

The \ac{gan} combines multiple suggestions from different \ac{gan} designs and 
was designed as a conditional~\cite{mirza2014} deep 
convolutional~\cite{radford2016} Wasserstein~\cite{arjovsky2017} \ac{gan} with 
gradient penalty~\cite{gulrajani2017} (cDC-WGAN-GP).  The design in terms of 
the intermediate image sizes is visualized in Fig.~\ref{fig:scGanStructure}.  
For the full design including all layers, consult Appendix~\ref{ap:gan}.

We opted for a design including a mixture between transposed, interpolation, 
and normal convolutional filter kernels, which prevented checkerboard 
artifacts and large patches.  The combination of interpolation layers and 
transposed convolutional layers lead to better images than each of the 
approaches alone (see also in Appendix~\ref{ap:scFailedGan} 
Fig.~\ref{fig:scFailedGan}) present in our previous 
approach~\cite{kaiser2022}.  The conditioning of the \ac{gan} was implemented 
as an embedding vector controlling the image label that we wished to 
synthesize.  We trained the \ac{gan} for 1000 epochs using the Wasserstein 
distance~\citep{arjovsky2017} which is considered to stabilize 
training~\citep{arjovsky2017a}.  Instead of the originally proposed weight 
clipping, we used a gradient penalty~\citep{gulrajani2017} of $\lambda=1$.  We 
used 10 critic iterations before updating the generator and a learning rate of 
$\alpha=3\cdot 10^{-5}$ for both networks.  The loss $L$ can be described as 
follows~\cite{gulrajani2017}:

\begin{equation}
    L = \Space{E}_{\tilde{x}\sim\Space{E}_D} D(\tilde{x} | y) - 
    \Space{E}_{x\sim\Space{E}_G} D(x | y) + \lambda ( || \nabla_{\hat{x}} D 
    (\hat{x}) ||_2 - 1)^2
    \label{eq:fun:ganDisc}
\end{equation}

with $\tilde{x}$ denoting the generator samples $G(z | y)$ and $\hat{x} = \epsilon 
x + (1-\epsilon) \tilde{x}$ with $\epsilon$ denoting a uniformly distributed 
random variable between 0 and 1~\cite{gulrajani2017}.

\begin{figure}[htpb]
\centering
\includegraphics[width=\textwidth]{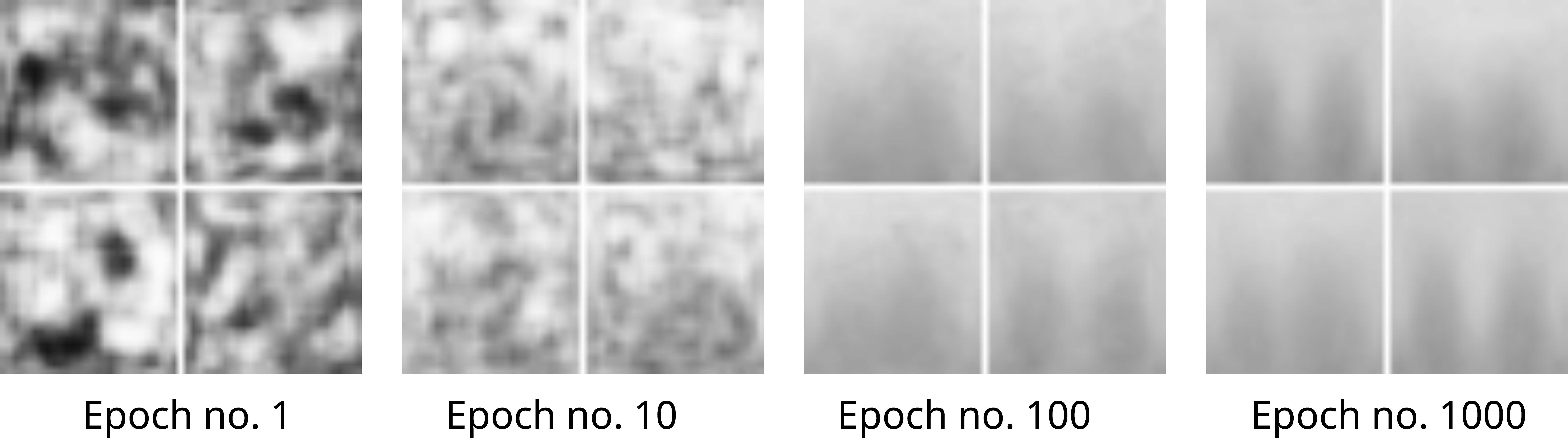} \\
\caption{Image development of the \ac{gan} generator during different stages 
of training visualized as a 2$\times$2 grid.}\label{fig:scGanTraining}
\end{figure}

\subsection{Image assessment}

We used \ac{ssimcc} as the basis for a metric to assess the similarity of the 
synthetic images to the clinical images and defined the 
$\mathrm{SSIM}_\mathrm{cc}$ for each \textit{synthetic} sample by using the 
minimum \ac{ssimcc} with respect to each \textit{clinical} sample of the same 
class $N$: 

\begin{equation}
    \mathrm{SSIM}_{\mathrm{cc},i} = \min\nolimits_{\forall n \in N} 
\mathrm{SSIM}(p_{i,\mathrm{synthetic}}, p_{n,\mathrm{clinical}})
\end{equation}

It has to be noted that the $\mathrm{SSIM}_\mathrm{cc}$ itself did not assess 
the quality of the synthetic images, but was rather designed to evaluate the 
similarity to the clinical images.  With this approach, we tried to quantify a 
``good'' data generator: The data should not be very similar to the original 
data (because then we could simply use the original data), but also not too 
different (because then they might not be a true representation of the 
underlying class anymore). ``Good'' images should not be ``too close'' to 1, 
but also not ``too low''.

\subsection{\Acs{cnn} Training}\label{subsec:scClassification}

\begin{figure}[htbp]
\centering
\includegraphics[width=0.8\textwidth]{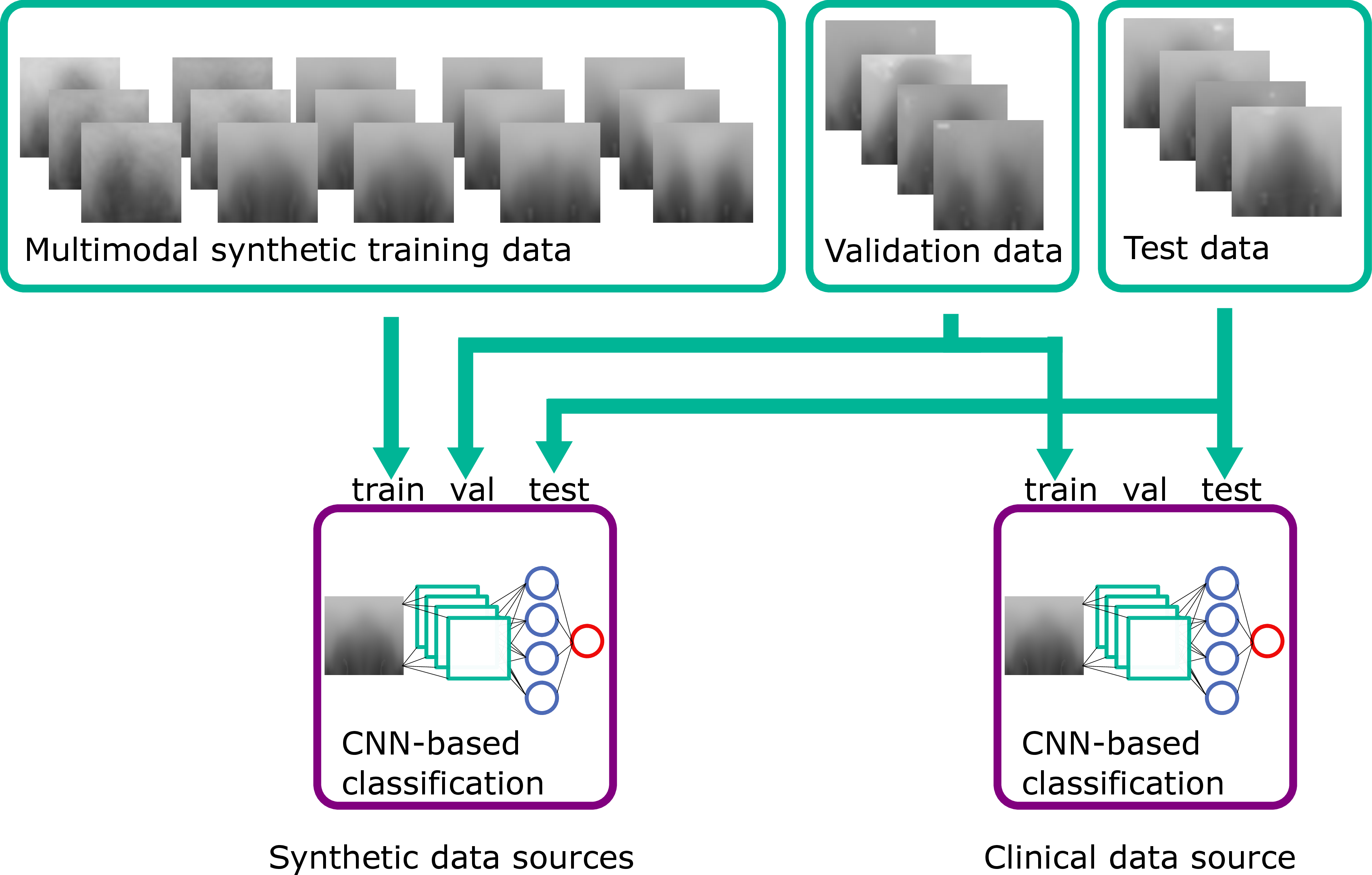}
\caption{Classification training using the synthetic data, the validation 
data, and the test set.  The \acs{cnn} classifier using clinical data uses the 
validation data as a training set.  Green: data, blue: violet: classification 
models.}%
\label{fig:sc:trainTest}
\end{figure}

Resnet18 was used as a classifier since it showed the best performance on this 
type of distance maps~\citep{schaufelberger2023}.  We used 
\texttt{pytorch}'s~\citep{paszke2019} publicly available, pre-trained Resnet18 
model and fine-tuned the weights during training.  During training, all images 
were reshaped to a size of 224$\times$224 to match the input size of Resnet18.  
We performed a different run of \ac{cnn} training on all seven combinations of 
the synthetic data.  The \ac{cnn} was trained only on synthetic data (except 
for the clinical scenario which was trained on clinical data for comparison).  
During training, we evaluated the model on both the (purely synthetic) 
training data and the (clinical) validation set (see also 
Fig.~\ref{fig:sc:trainTest}).  The best-performing network was chosen 
according to the maximum F1-score on the validation set.  The test set was 
never touched during training and only evaluated in a final run after 
training.  

\begin{figure}[htbp]
\centering
\includegraphics[width=0.8\textwidth]{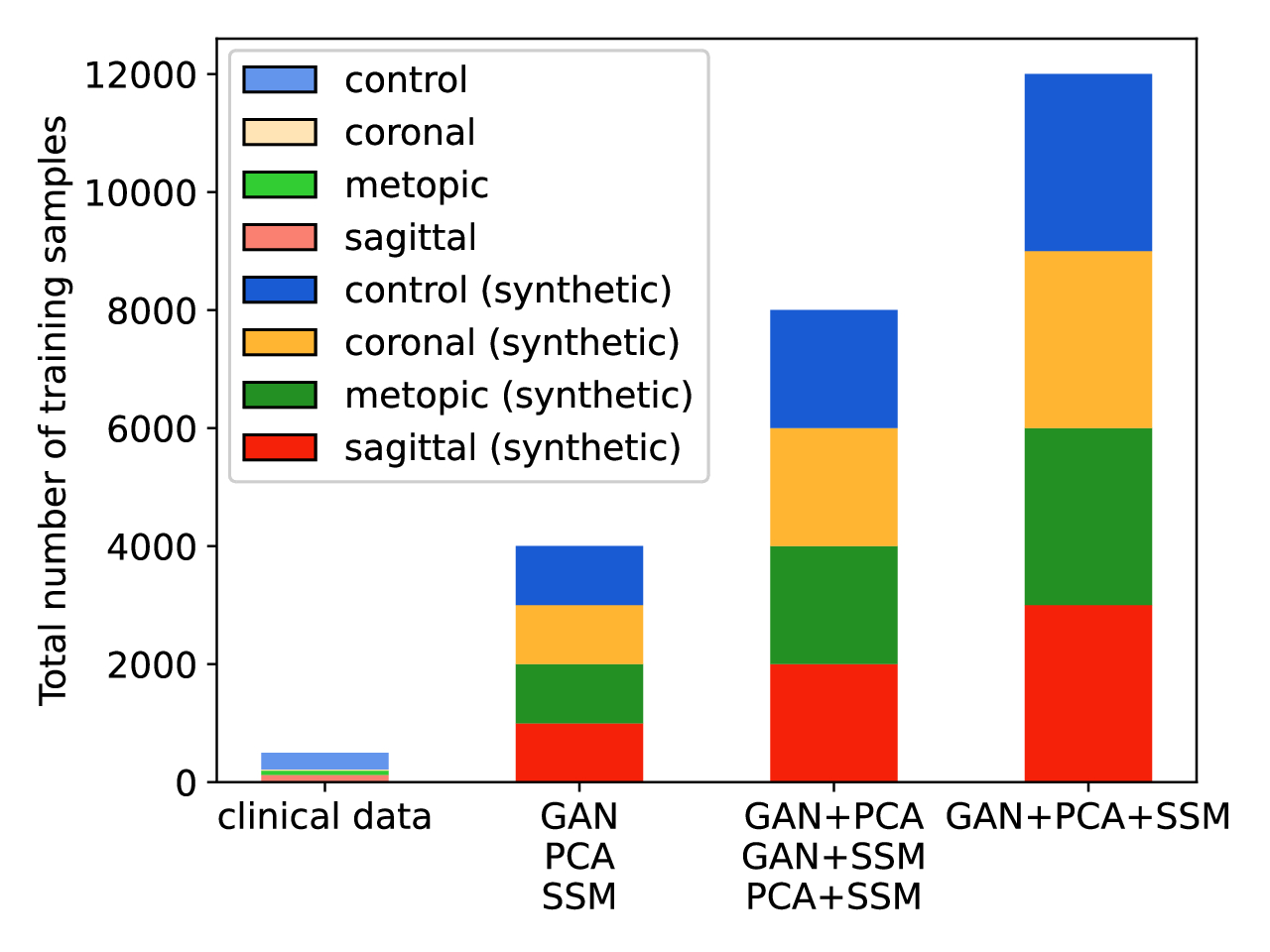}
\caption{Number of training samples in each classification scenario.  The 
clinical scenario has less than 500 samples while all synthetic scenarios have 
at 4000, 8000, or 12000 samples.}%
\label{fig:sc:numberSamples}
\end{figure}

When multiple data sources were used, the models had a different number of 
training samples (see Fig.~\ref{fig:sc:numberSamples}) and all 
synthetically-trained models were trained for 50 epochs.  Convergence was 
achieved usually already during the first ten epochs, indicating that there 
was sufficient training material for each model.  We used Adam optimizer, 
cross entropy loss, a batch size of 32 with a learning rate of 
$1\cdot10^{-4}$, weight decay of $0.63$ after each 5 epochs.  To evaluate the 
synthetically-trained models against a clinically trained model, we 
additionally employed one \ac{cnn} trained on clinical data,  trained with the 
same parameters except a higher learning rate of $1\cdot10^{-3}$.  

We used the following types of data augmentation during training: Adding 
random pixel noise (with $\sigma=1/255$), adding a random intensity (with 
$\sigma=5/255$) across all pixels, horizontal flipping, and shifting images 
left or right (with $\sigma=12.44\,\mathrm{pixels}$).  All those types of data 
augmentation corresponded to real-world patient and scanning modifications: 
Pixel noise corresponded to scanning and resolution errors, adding a constant 
intensity was equal to a re-scaling of the patient's head, horizontal flipping 
corresponded to the patient as if they were mirrored in real life, and 
shifting the image horizontally modeled an alignment error in which the 
patient effectively turns their head $20^\circ$ left or right during 
recording.  

All the clinical 2D data, the \ac{gan}, and the statistical models were made 
publicly 
available\footnote{\href{https://github.com/KIT-IBT/craniosource-gan-pca-ssm}{https://github.com/KIT-IBT/craniosource-gan-pca-ssm}}.  
We included a script to create synthetic samples for all three image 
modalities to allow users to create a large number of samples.  The synthetic 
and clinical samples of this study are available on 
Zenodo~\citep{schaufelberger2023a}.

\section{Results}

\subsection{Image evaluation}

\begin{figure}
\centering
\includegraphics[width=0.8\textwidth]{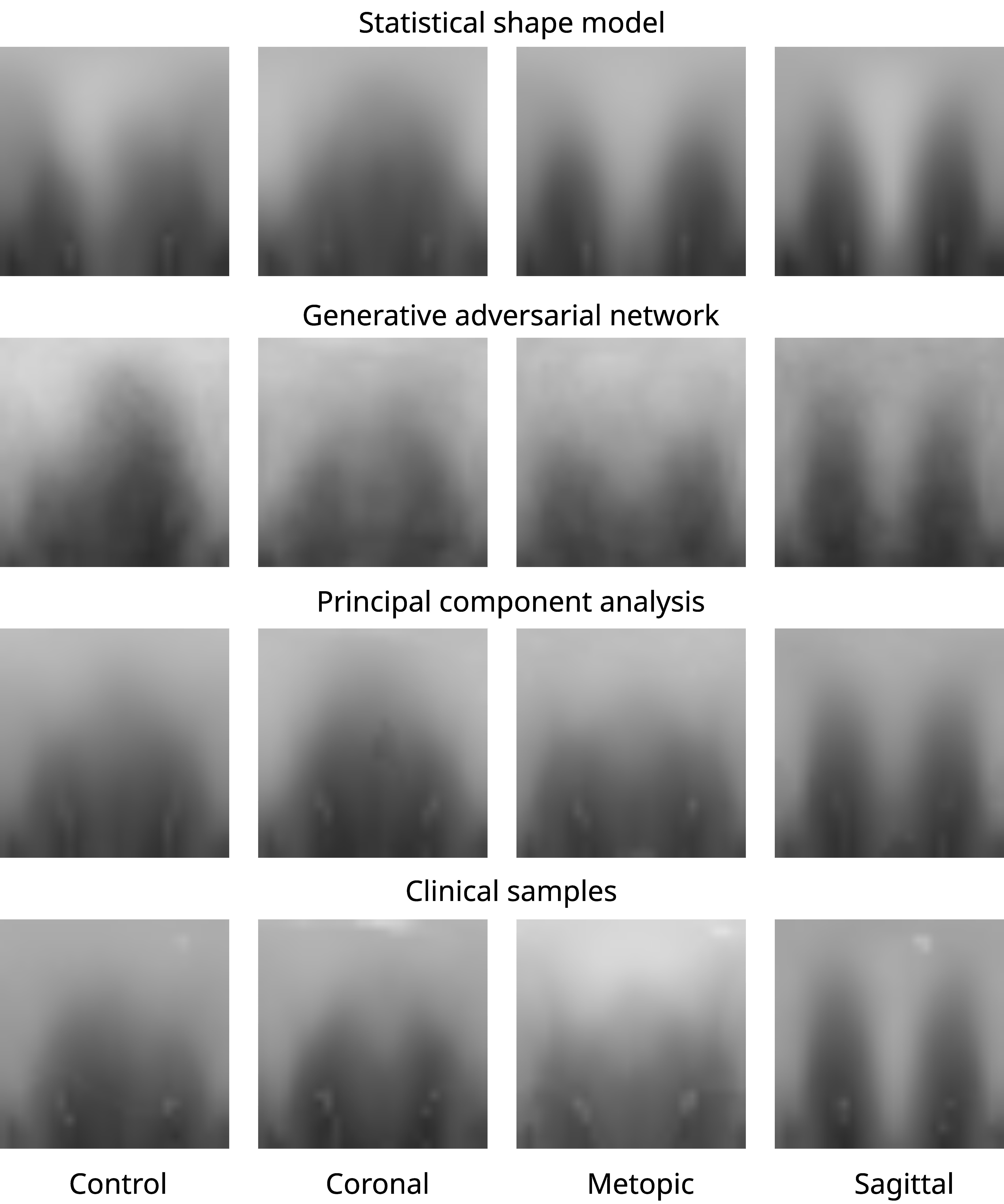}
\caption{Images of all three data modalities and clinical samples.  From top to bottom the image modalities: \ac{ssm}, \ac{gan}, \ac{pca}, clinical. From 
left to right the four classes: Control, coronal, metopic, 
sagittal.}\label{fig:scImgs}
\end{figure}

Fig.~\ref{fig:scImgs} shows image of each of the different data synthesis 
types compared with the clinical images.  From a qualitative, visual 
examination, the synthetic images had similar color gradients, shapes, and 
intensities as the clinical images.  \Ac{gan} images appeared slightly noisier 
than the other images and did not show the left and right ear visible in the 
other images.

\begin{figure}[htbp]
\centering
\includegraphics[width=\textwidth]{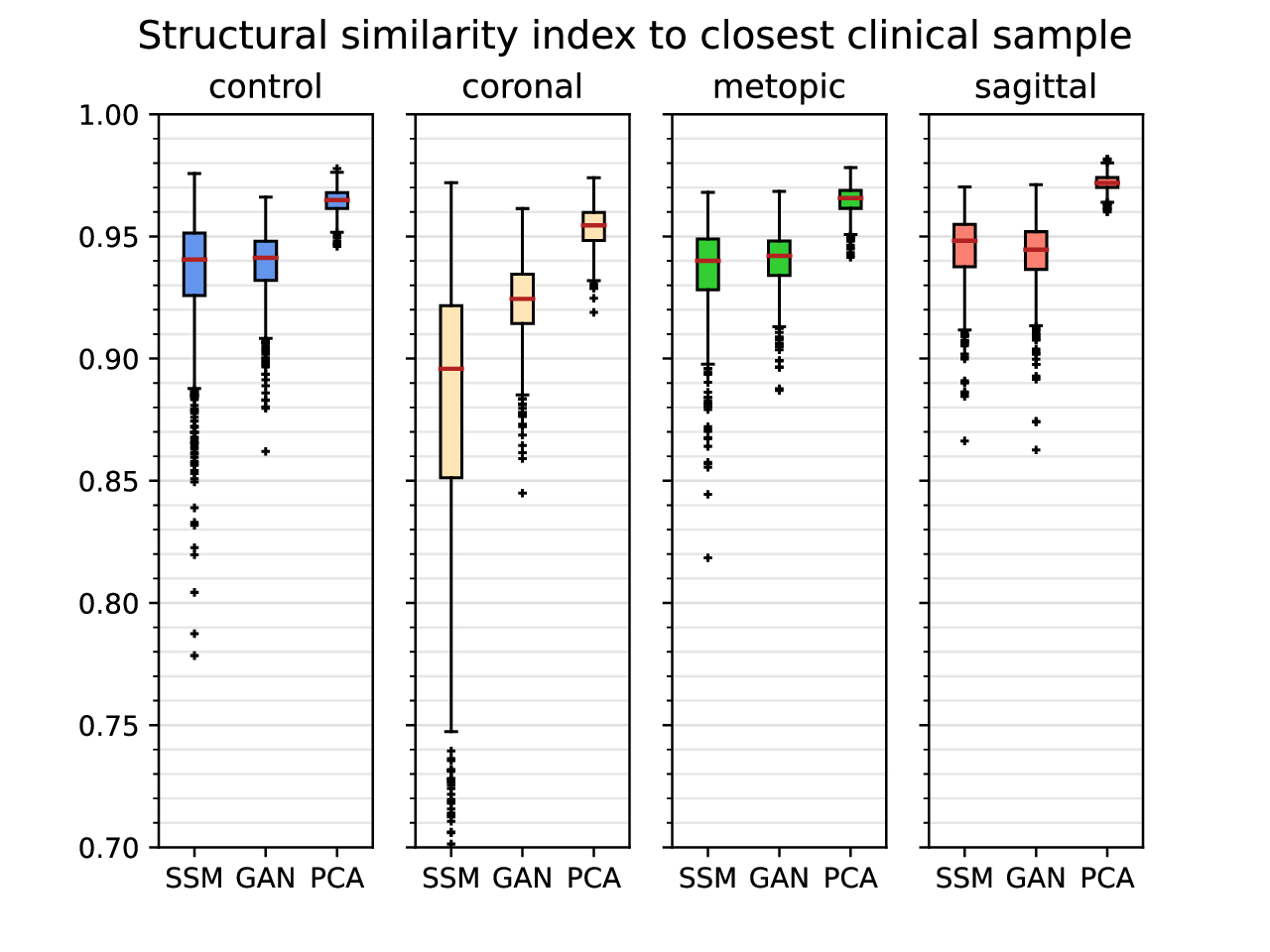}
\caption{Boxplots of $\mathrm{SSIM}_\mathrm{cc}$ (\acl{ssim} to closest 
clinical sample) of each class for each of the synthetic data generators.}%
\label{fig:scSSIM}
\end{figure}

From the quantitative comparison (see Fig.~\ref{fig:scSSIM}), ordinary 
\ac{pca} images were substantially and consistently more similar to the 
clinical images than the other two modalities (differences of the medians 
larger than 0.02), while \ac{ssm} and \ac{gan} images were less similar, with 
the \ac{ssm} images being the most dissimilar for the coronal class.

\subsection{Classification results}

All comparison presented here were carried out on the untouched test set.  
According to the classification results for the synthetic training in 
Tab.~\ref{tab:scCompSynth}, the \ac{ssm} was the best single source of 
synthetic data with an F1-score higher than 0.85.  All combinations of 
synthetic models showed F1-scores higher than 0.8.  The classifier on the 
clinical data scored an accuracy above 0.96, but was surpassed by the 
combination of \ac{gan} and \ac{ssm}. F1-score was highest for the clinical 
classification (0.9533), but the combination of \ac{ssm} and \ac{gan} scored a 
very close F1-score (0.9518).  Including a second data source always improved 
the F1-score compared to a model with a single data source (adding \ac{pca} to 
\ac{gan} by 0.29, adding \ac{ssm} to \ac{pca} by 0.16, adding \ac{ssm} to 
\ac{gan} by 0.1).

\begin{table}[htbp]
\caption{\acs{cnn}-Classification comparison on the test set trained on 
different synthetic data sources.  Boldface: best results among the data 
source.}%
\label{tab:scCompSynth}
\centering
\begin{tabular}{lrr}
    \toprule
    Synthetic data source    & Accuracy        & F1-score        \\
    \midrule
    GAN          & 0.4274  & 0.4930 \\
    PCA          & 0.7581  & 0.6997 \\
    SSM          & 0.9153  & 0.8547 \\
    GAN-PCA      & 0.8508  & 0.7823 \\
    GAN-SSM      & \textbf{0.9677}  & \textbf{0.9518} \\
    PCA-SSM      & 0.9153  & 0.8595 \\
    GAN-PCA-SSM  & 0.9597  & 0.9445 \\
    \midrule
    Clinical            & 0.9637  & 0.9533 \\
    \bottomrule
\end{tabular}
\end{table}

\section{Discussion}\label{sec:scDiscussion}

Without being trained on a single clinical sample, the \ac{cnn} trained from 
the combination of the \ac{ssm} and the \ac{gan} was able to correctly 
classify 95\,\% of the data.  Classification performance on the synthetic data 
proved to be equal to or even slightly better than training on the clinical 
data, at least for the data generated using the \ac{ssm} and the \ac{gan} (and 
optionally also \ac{pca}).  This suggests that certain combinations of 
synthetic data might be indeed sufficient for a classification algorithm to 
distinguish between types of craniosynostosis.  Compared with classification 
results from other works, the purely synthetic-data-based classification 
performs in a similar range and sometimes even better than other approaches on 
clinical 
data~\citep{dejong2020,schaufelberger2023,schaufelberger2022a,agarwal2021,mendoza2014}.

The \ac{ssm} appeared to be the data source contributing the most to the 
improvement of the classifier: Not only did it score highest among the unique 
data sources, but it was also present in the highest scoring classification 
approaches.  One reason for this might be that it was also the least similar 
data source for most of the classes.  Due to the inherent modeling of the 
geometric shape in 3D, the created 2D distance maps are always created from 3D 
samples, while \ac{pca} and the \ac{gan} could, in theory, create 2D images 
which do not correspond to a 3D shape.  In contrast, the \ac{gan}-based 
classifiers only showed a good classification performance when combined with a 
different data modality and its synthesized images seemed to show less 
pronounced visual features than the other two modalities.  However, the 
\ac{ssimcc} based metric shows no substantial difference between the \ac{gan} 
images and the other two modalities.  However, one possible reason might be 
that the \ac{gan} learned features of multiple classes and the images might 
still contain features which are derived from images from other classes.  The 
\ac{pca} images were neither required, nor detrimental for a good 
classification performance.  According to the \ac{ssimcc}, the \ac{pca} images 
were the most similar images to its clinical counterparts.  

Overall, a combination of different data modalities seemed to be the key 
element for achieving a good classification performance.  Both \ac{ssm} and 
\ac{pca} model the data according to a Gaussian distribution, while the 
\ac{gan} uses an unrestricted distribution model.  The different properties of 
modeling the underlying statistical distribution of a Gaussian distribution 
(\acp{ssm} and \ac{pca}) on the one hand, and without an assumed distribution 
(\ac{gan}) on the other hand might have lead to a compensation of their 
respective disadvantage increasing overall performance for the combinations.
One limitation of this study is the small dataset.  As the clinical 
classification uses the same dataset for training and validation, this might 
make it prone to overfitting.  However, the resulting classification metrics 
achieved in this study were similar to a classification study on clinical data 
alone~\citep{schaufelberger2023} which suggests that over-fitting has not been 
an issue.

\section{Conclusion}\label{sec:scConclusion}

We showed that it is possible to train a classifier for different types of 
craniosynostosis based solely on artificial data synthesized by a \ac{ssm}, 
\ac{pca}, and a \ac{gan}.  Without having seen any clinical samples, a 
\ac{cnn} was able to classify four types of head deformities with an F1-score 
higher than 0.95 and performed comparable to a classifier trained on clinical 
data.  The key component in achieving good classification results was using 
multiple, but different data generation models.  Overall, the \ac{ssm} was the 
data source contributing most to the classification performance.  For the 
\ac{gan}, using a small image size and alternating between transposed 
convolutions and interpolations were identified as key elements for suitable 
image generation.  The datasets and generators were made publicly available 
along with this work.  We showed that clinical data is not required for the 
classification of craniosynostosis paving the way into cost-effective usage of 
synthetic data for automated diagnosis systems.

\bibliography{combined}

\newpage 
\appendix

\section{Generative adversarial network structure}\label{ap:gan}

This is the \ac{gan} structure of generator and discriminator employed for the 
creation of the synthetic data as an output of according to models' 
\texttt{\_\_str\_\_} method called via \texttt{print(model)}.

{\footnotesize
\begin{verbatim}
Generator28(
  (embed): Embedding(4, 100)
  (gen): Sequential(
    (0): Sequential(
      (0): ConvTranspose2d(200, 256, kernel_size=(5, 5), stride=(1, 1), 
                bias=False)
      (1): BatchNorm2d(256, eps=1e-05, momentum=0.1, affine=True, 
                track_running_stats=True)
      (2): ReLU(inplace=True)
    )
    (1): Sequential(
      (0): Interpolate(size=(8, 8),bilinear,align_corners=True)
      (1): BatchNorm2d(256, eps=1e-05, momentum=0.1, affine=True, 
                track_running_stats=True)
      (2): ReLU(inplace=True)
    )
    (2): Sequential(
      (0): Conv2d(256, 128, kernel_size=(3, 3), stride=(1, 1), padding=(1, 1), 
                bias=False)
      (1): BatchNorm2d(128, eps=1e-05, momentum=0.1, affine=True, 
                track_running_stats=True)
      (2): ReLU(inplace=True)
    )
    (3): Sequential(
      (0): Interpolate(size=(15, 15),bilinear,align_corners=True)
      (1): BatchNorm2d(128, eps=1e-05, momentum=0.1, affine=True, 
                track_running_stats=True)
      (2): ReLU(inplace=True)
    )
    (4): Sequential(
      (0): ConvTranspose2d(128, 128, kernel_size=(3, 3), stride=(1, 1), 
                bias=False)
      (1): BatchNorm2d(128, eps=1e-05, momentum=0.1, affine=True, 
                track_running_stats=True)
      (2): ReLU(inplace=True)
    )
    (5): Sequential(
      (0): Interpolate(size=(30, 30),bilinear,align_corners=True)
      (1): BatchNorm2d(128, eps=1e-05, momentum=0.1, affine=True, 
                track_running_stats=True)
      (2): ReLU(inplace=True)
    )
    (6): Conv2d(128, 1, kernel_size=(3, 3), stride=(1, 1), 
                bias=False)
    (7): Tanh()
  )
)

Discriminator28(
  (net): Sequential(
    (0): Sequential(
      (0): Conv2d(2, 32, kernel_size=(4, 4), stride=(2, 2), padding=(1, 1), 
                bias=False)
      (1): InstanceNorm2d(32, eps=1e-05, momentum=0.1, affine=True, 
                track_running_stats=False)
      (2): LeakyReLU(negative_slope=0.2)
    )
    (1): Sequential(
      (0): Conv2d(32, 128, kernel_size=(4, 4), stride=(2, 2), padding=(1, 1), 
                bias=False)
      (1): InstanceNorm2d(128, eps=1e-05, momentum=0.1, affine=True, 
                track_running_stats=False)
      (2): LeakyReLU(negative_slope=0.2)
    )
    (2): Sequential(
      (0): Conv2d(128, 256, kernel_size=(5, 5), stride=(2, 2), padding=(1, 1), 
                bias=False)
      (1): InstanceNorm2d(256, eps=1e-05, momentum=0.1, affine=True, 
                track_running_stats=False)
      (2): LeakyReLU(negative_slope=0.2)
    )
    (3): Conv2d(256, 1, kernel_size=(3, 3), stride=(1, 1))
  )
  (embed): Embedding(4, 784)
)
\end{verbatim}
}

\section{Failed GAN attempts}\label{ap:scFailedGan}

\begin{figure}[hbtp]
\centering
\includegraphics[width=\columnwidth]{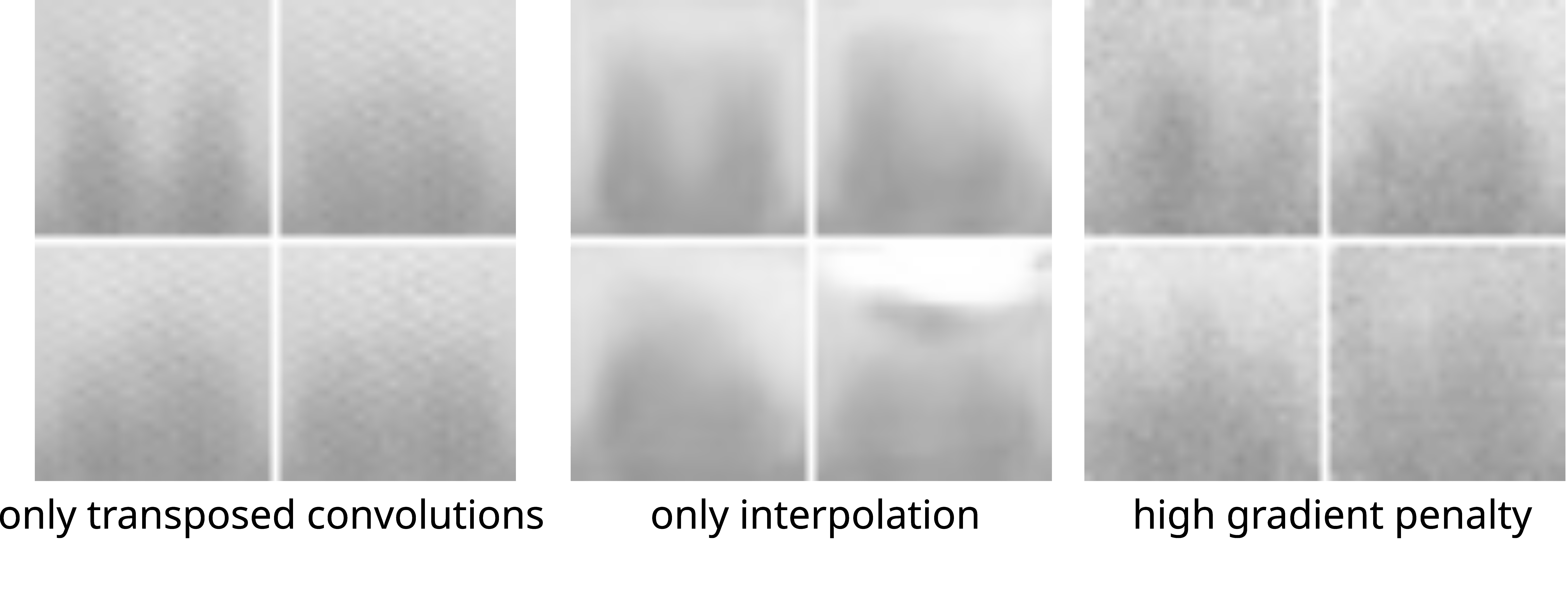}
\caption{Failed \ac{gan} images arranged in a 2$\times$2 grid with artifacts 
arising from poor network design or bad training conditions. From left to 
right: Deconvolution artifacts, interpolation artifacts, and noise 
artifacts.}\label{fig:scFailedGan}
\end{figure}

We show artifacts arising from only using transposed convolutional layers 
\linebreak
(\texttt{ConvTranspose2d}), using only up-scaling interpolation layers 
(\texttt{Interpolate}), or from large gradient penalties which prohibits 
training in Fig.~\ref{fig:scFailedGan}.

\end{document}